\documentclass[preprint,12pt]{elsarticle}

\usepackage{color}
\usepackage{amssymb}
\usepackage{graphicx,color}
\usepackage[usenames,dvipsnames]{xcolor}
\usepackage[section]{placeins}
\usepackage{amsmath}
\usepackage{subfigure}
\usepackage[normalem]{ulem}
\usepackage{booktabs}
\usepackage{xcolor}
\usepackage{epsfig,graphics,color,graphicx,amsmath}

\colorlet{darkgreen}{green!50!black}
\colorlet{brightyellow}{yellow!75!red}
\colorlet{orange}{red!50!yellow}
\colorlet{darkblue}{blue!60!black}
\colorlet{darkred}{red!80!black}

\usepackage{soul}

\usepackage{mathtools}
\usepackage{mathrsfs}
\usepackage{bm}
\usepackage{relsize}
\usepackage{indentfirst}

\usepackage{xcolor}

\journal{Physics Letters B}
\begin{document}
\begin{frontmatter}
\title{Solving the three-body bound-state  Bethe-Salpeter equation in Minkowski space}
\author{E. Ydrefors$^a$} 
\author{J.H. Alvarenga Nogueira$^{a,b}$}
\author{V.A. Karmanov$^c$}
\author{T. Frederico$^a$}
\address{$^a$Instituto Tecnol\'ogico de Aeron\'autica,  DCTA, 
12228-900 S\~ao Jos\'e dos Campos,~Brazil}
\address{$^b$Dipartimento di Fisica, Universit\`a di Roma ``La Sapienza" \\
INFN, Sezione di Roma ``La Sapienza"
Piazzale A. Moro 5 - 00187 Roma, Italy}
\address{$^c$Lebedev Physical Institute, Leninsky Prospekt 53, 119991 Moscow, Russia}
\date{\today}

\begin{abstract}
The scalar three-body Bethe-Salpeter equation, with zero-range interaction, is solved in \\ Minkowski space  by direct integration of the four-dimensional integral equation. The singularities appearing in the propagators are treated properly by standard analytical and numerical methods, without relying on any ansatz or assumption. The results for the binding energies and transverse amplitudes are compared with the results computed in Euclidean space. A fair agreement between the calculations is found.
\end{abstract}
\end{frontmatter}

\section{Introduction}\label{Sec:intr}
The Bethe-Salpeter (BS) equation \cite{BS}, formally defined in the Minkowski space, is an efficient tool to study relativistic systems in the non-perturbative regime. One of the commonest methods to solve the BS equation numerically is to perform an analytic continuation to the complex plane, through the Wick rotation 
\cite{W_54}, into the Euclidean space. After this transformation, the equation turns to be non-singular as the singularities are moved from the integration line (real axis) to the complex plane. This is the widely used method of finding the binding energies, especially for the two-body BS 
equation. However, though certain quantities such as binding energies and transverse amplitudes exactly coincide with those determined by the Minkowski BS equation, the Euclidean BS amplitude 
is not the physical one and does not give direct access to most of the dynamical observables.  For example,  the  electromagnetic transition form factor, 
associated with the breakup of a two-body bound state, can be computed in Minkowski space  in the whole kinematical region including the final state interaction \cite{Carbonell:2015awa}, while
this task has not yet been accomplished with Euclidean space calculations.
For general purposes one needs the Minkowski BS amplitude. One successful way of solving the BS equation fully in Minkowski space is by 
looking for the solution in the form of the Nakanishi integral representation \cite{KW} 
combined with the light-front projection \cite{bs1,FrePRD14}. Another alternative, without relying on any ansatz for the solution, is by direct integration of the 
BS equation poles and singularities \cite{ck2b}, present in the propagators and amplitude. Although the direct integration method is practicable, it is much more demanding numerically than solving the problem in the Euclidean space. Nevertheless, all these methods were successfully applied to two-body systems.  

The relativistic three-body systems are extremely interesting, with widespread worthwhile applications, but also more challenging. Most of the extensive researches in the three-body context were carried out in the  contact interaction framework, which, in spite of its simplification, remains to be rather instructive. The zero-range interaction BS and light-front (LF) equations for the bound state of three scalar particles, by means of the Faddeev decomposition, were derived  in Ref.~\cite{tobias1}. The LF equation was firstly solved in Ref.~\cite{tobias1} and its solution was re-analyzed in Ref.~\cite{ck3b}. The three-body BS equation \cite{tobias1} was also recently solved, for the first time, in Euclidean space \cite{ey3b}. Although the LF equation is fully defined in Minkowski space, it gives access only to the valence component, which is far from enough for relativistic calculations.  According to \cite{ey3b}, the contributions from higher-Fock components are remarkable and cannot be neglected,  as already anticipated in
\cite{Karmanov:2008bx}.
This is an important motivation for going beyond the approaches based on the valence component of the LF wave function introduced in \cite{tobias1}. To obtain observables, considering the many-body components beyond the valence consistently is critical to solve the four-dimensional equation fully in Minkowski space. Thereby the aim of the present work is to solve the bosonic three-body BS equation \cite{tobias1} directly in the Minkowski space, without relying on any anzats or three-dimensional reduction. Finding the solution of three-body equation in the form of the Nakanishi integral representation is a work in progress. Extending the ``arsenal" of the methods is useful for comprehending the BS equation in more realistic cases and this is a first step dealing with the three-body equation in Minkowski space. 

The rest of this paper is organized as follows. In Sec.~\ref{Sec:BSE}, we transform the BS equation to a partially non-singular form. The expressions for the transverse amplitudes are derived in Sec.~\ref{Sec:TA}. Sec.~\ref{Sec:Res} presents the numerical results for the binding energies, BS amplitudes and transverse amplitudes. Finally, in Sec.~\ref{Sec:Concl} we draw our conclusions.

\section{Bethe-Salpeter equation}\label{Sec:BSE}
We  consider  the system of three scalar bosons with equal constituent masses $m$ with zero-range interaction. The Faddeev component of the vertex function $v(q,p)$ complies with a single integral equation given by \cite{tobias1}
\begin{equation}
v(q,p)=2iF(M_{12})\int \frac{d^4 k}{(2\pi)^4}\frac{i}{[k^2-m^2+i\epsilon]}\frac{i}{[(p-q-k)^2-m^2+i\epsilon]}v(k,p).
\label{Eq:BSE}
\end{equation}
Due to the two-body zero-range interaction the vertex function $v(q,p)$ depends only on the total four momentum $p$ and the four momentum of the spectator particle $q$. Furthermore, $F(M_{12})$ denotes the two-body scattering amplitude namely the zero-range interaction kernel, and it reads
\begin{equation}
\mathcal{F}(M_{12})=\left\{\begin{aligned}&\frac{1}{\frac{1}{16\pi^2 y}\log\frac{1+y}{1-y}-\frac{1}{16\pi m a}} \quad ; \: M_{12}^2\leq 0, \\
&\frac{1}{\frac{1}{8\pi^2 y'}\arctan y'-\frac{1}{16\pi m a}} \quad ; \: 0\leq M_{12}^2\leq 4m^2,
\\& \frac{1}{\frac{y''}{16\pi^2}\log\frac{1+y''}{1-y''}-\frac{1}{16\pi m a}-i\frac{y''}{16\pi}} \quad ; \: M_{12}^2\geq 4m^2,
\end{aligned}\right.
\label{Eq:F_amp}
\end{equation}
with
\begin{equation}
y=\frac{\sqrt{-M_{12}^2}}{\sqrt{4m^2-M_{12}^2}},
\quad
y'=\frac{M_{12}}{\sqrt{4m^2-M_{12}^2}},
\quad
y''=\frac{\sqrt{M_{12}^2-4m^2}}{M_{12}}.
\end{equation}
The squared effective mass of the two-body subsystem (not including the spectator) takes the form $M_{12}^2=(p-q)^2$ and $a$ denotes the scattering length, which is the renormalization parameter used to regularize the bubble diagram.

The Eq.~(\ref{Eq:BSE}) constitutes a singular integral equation and must be rewritten in a non-singular form before it can be solved numerically. As mentioned, for this aim, in \cite{ey3b} we transformed Eq.~(\ref{Eq:BSE}) into the Euclidean space. In the Minkowski space,  the strongest singularities (the pole singularities) are present in the propagators. For their treatment, in this paper we use the direct method introduced in \cite{ck2b}. The first step is to represent the propagator $[k^2-m^2+i\epsilon]^{-1}$ as follows
\begin{equation}
\frac{1}{k^2-m^2+i\epsilon}=\frac{1}{k_0^2-k^2_v-m^2+i\epsilon}=PV\frac{1}{k_0^2-\varepsilon_k^2}-\frac{i\pi}{2\varepsilon_k}[\delta(k_0-\varepsilon_k)+\delta(k_0+\varepsilon_k)],
\end{equation} 
where $\varepsilon_k=\sqrt{k_v^2+m^2}$ and $k_v=|\vec{k}|$.  Then we eliminate the singularities of the integrands in the form of $PV\int\ldots\frac{dk_0}{k_0^2-\varepsilon_k^2}$, exploiting the following identities 
\begin{equation}
PV \int_{-\infty}^0 \frac{dk_0}{k^2_0-\varepsilon_k^2}=PV \int_{0}^{\infty} \frac{dk_0}{k^2_0-\varepsilon_k^2}=0,
\end{equation}
with appropriate coefficients, to subtract the kernel at the singular point.
After subtracting, the PV  (principal value) integrals become smooth and the symbol PV is dropped out.  

As for the second propagator in \eqref{Eq:BSE}, we integrate it,  in the c.m.-frame, i.e $\vec{p}=0$, over $z=\cos\left(\frac{\vec{k}\cdot\vec{q}}{k_v q_v}\right)$ (and multiply by $2\pi$ from the azimuthal angle integration). The result reads:
\begin{eqnarray}\label{Eq:PI}
\Pi(q_0,q_v,k_0,k_v)   =\int\frac{idzd\varphi}{[(p-q-k)^2-m^2+i\epsilon]}= \frac{i \pi}{q_v k_v} \left\{ \log \left| \frac{(\eta+1)}{(\eta-1)} \right| - i \pi I(\eta)   \right\},
\end{eqnarray}
with
\begin{equation}\label{Eq:eta}
I(\eta)=\left\{
\begin{array}{lcrcl}
1  & {\rm if} & \mid\eta\mid &\leq& 1 \cr
0  & {\rm if} & \mid\eta\mid &> & 1
\end{array}\right.,
\end{equation}
and
\begin{equation}
\eta =   \frac{(M_3 - q_0 - k_0)^2 - k_v^2 - q_v^2 - m^2}{2q_v k_v}.
\end{equation}
The $\log$-singularity in \eqref{Eq:PI} can be then integrated by standard numerical methods.

After these transformations, the equation \eqref{Eq:BSE} for the Faddeev component of the three-body vertex function in the rest frame obtains the following form 
\begin{eqnarray}
v(q_0,q_v)  
&=& \frac{\mathcal{F}(M_{12})}{(2\pi)^4} \int_0^{\infty} k^2_v dk_v\left\lbrace   \frac{2\pi i}{2 \varepsilon_k}  \left[\Pi(q_0,q_v;\varepsilon_k,k_v) v(\varepsilon_k,k_v)  + \Pi(q_0,q_v;-\varepsilon_k,k_v)  v(-\varepsilon_k,k_v)\right]\right.
\cr
   &-&2 \int^0_{-\infty} dk_0 \left[ \frac{ \Pi(q_0,q_v;k_0,k_v) v(k_0,k_v)  - \Pi(q_0,q_v;-\varepsilon_k,k_v)  v(-\varepsilon_k,k_v)}{  {k}_0^2-\varepsilon_k^2 }\right]  \cr
&-&  \left.   2 \int_0^{\infty} dk_0 \left[ \frac{ \Pi(q_0,q_v;k_0,k_v) v(k_0,k_v)  - \Pi(q_0,q_v;\varepsilon_k,k_v)  v(\varepsilon_k,k_v)}{  {k}_0^2-\varepsilon_k^2 }\right]\right\rbrace,
                   \label{Eq:v}
\end{eqnarray}
The integrand here, in contrast to the integrand of (\ref{Eq:BSE}), is not singular anymore at $k_0=\pm\varepsilon_k$.
Instead, the kernel $\Pi$, defined in Eq.~\eqref{Eq:PI}, has logarithmic singularities at $\eta=\pm 1$. For fixed values of $q_0$, $q_v$ and $k_v$, the singular points for $\Pi(q_0,q_v,k_0,k_v)$ v.s. $k_0$ are
\begin{eqnarray}
k_0&=&(M_3-q_0)+\sqrt{m^2+(k_v\pm q_v)^2},
\nonumber\\
k_0&=&(M_3-q_0)-\sqrt{m^2+(k_v\pm q_v)^2}.
\label{Eq:sing_k0}
\end{eqnarray}

Similarly, the singular points of the kernel $\Pi(q_0,q_v,\pm\varepsilon_k,k_v)$ v.s. $k_v$ are given by
\begin{equation}
k_v=\frac{\pm \sqrt{M_{12}^2(M_{12}^2+q_v^2)(M_{12}^2-4m^2)}\pm q_v M_{12}^2}{2M_{12}^2},
\label{Eq:sing_k}
\end{equation}
where $M_{12}^2=(M_3-q_0)^2-q^2_v$.
The expression under the square root is non-negative if
$$
M_{12}^2 \geq 4m^2 \quad \mbox{or} \quad M_{12}^2 \leq 0.
$$
Consequently, real singular points $k_v$ exist if
\begin{equation}
q_0<M_3-\sqrt{q^2_v+4m^2}\quad \text{or}\quad M_3-q_v<q_0<M_3+q_v \quad\text{or}\quad q_0>M_3+\sqrt{q^2_v+4m^2}.
\end{equation}
The transition points in the variable $q_0$ between two regimes (with and without singularities v.s. $k_v$)  are thus
\begin{equation}
\label{Eq:peaks}
\begin{aligned}
q^{(1)}_0=&M_3-\sqrt{q^2_v+4m^2}, \\
q^{(2)}_0=&M_3-q_v, \\
q^{(3)}_0=&M_3+q_v, \\
q^{(4)}_0=&M_3+\sqrt{q^2_v+4m^2},
\end{aligned}
\end{equation}
with $q^{(1)}_0<q^{(2)}_0< q^{(3)}_0< q^{(4)}_0$, and they coincide with the transition points of the two-body amplitude $F(M_{12})$, see Eq.~\eqref{Eq:F_amp}.  These inequalities define five intervals of the variable $q_0$: between and outside these points. Knowing these intervals is extremely useful, as it allows to perform a much cleverer treatment of the weakly logarithmic singularities numerically.

\section{Transverse amplitudes}\label{Sec:TA}

The Minkowski vertex function $v(q_0,q_v)$ cannot be directly compared with the corresponding Euclidean one. However, in the BS amplitude, one can instead of $k=(k_0,k_v)$ introduce the light-front variables $k=(k_-,k_+,\vec{k}_\perp)$, where $k_{\mp}=k_0 \mp k_z$ and $\vec{k}_\perp=(k_x,k_y)$. The transverse amplitudes -- double integrals of the Minkowski BS amplitude  over $k_+$ and $k_-$, and of the corresponding Euclidean amplitude over  $k_0$, $k_z$, -- are then the same (up to a Jacobian).  Below we will calculate the transverse amplitudes using  the Minkowski BS amplitude, so we can compare the solution of this paper with the one found previously through the BS equation solved in Euclidean space \cite{ey3b}. In this section we will perform the integrations over  $k_+$ and $k_-$.

The BS amplitude can be written in terms of the three vertex components as
\begin{equation}
i\Phi_M(k_1,k_2,k_3;p)=i^3\frac{v_M(k_1)+v_M(k_2)+v_M(k_3)}{(k^2_1-m^2+i\epsilon)(k^2_2-m^2+i\epsilon)(k^2_3-m^2+i\epsilon)},
\end{equation}
where the four-momenta obeys the relation
\begin{equation}
k_1+k_2+k_3=p.
\end{equation}

We subsequently define the transverse amplitude as the integral
\begin{equation}
\begin{aligned}
L(\vec{k}_{1\perp}, \vec{k}_{2\perp})&=L_1(\vec{k}_{1{\perp}}, \vec{k}_{2{\perp}})+L_2(\vec{k}_{1{\perp}}, \vec{k}_{2{\perp}})+L_3(\vec{k}_{1{\perp}}, \vec{k}_{2{\perp}})=\\
& \int_{-\infty}^{\infty}dk_{10} \int_{-\infty}^{\infty} dk_{1z} \int_{-\infty}^{\infty} dk_{20} \int_{-\infty}^{\infty} dk_{2z}\;i\Phi_M(k_{10},k_{1z},k_{20},k_{2z}; \vec{k}_{1\perp}, \vec{k}_{2\perp}).
\end{aligned}
\end{equation}

As for the equal masses case, we can deal with one of the components, which is given by
\begin{equation}
L_1(\vec{k}_{1\perp}, \vec{k}_{2\perp})=i\int_{-\infty}^{\infty}dk_{10}\int_{-\infty}^{\infty}dk_{1z}\;\frac{v_M(k_{10},k_{1v})}{k_1^2-m_1^2+i\epsilon}\chi(k_{10},k_{1z},k_{20},k_{2z}),
\end{equation}
where 
\begin{equation}
\label{Eq:chi}
\chi(k_{10},k_{1z},k_{20},k_{2z})=i^2\int\frac{d^2k_2}{(k^2_2-m_2^2+i\epsilon)[(p'-k_2)^2-m_3^2+i\epsilon]}.
\end{equation}
In Eq.~\eqref{Eq:chi} we have used
\begin{equation}
k_i=(k_{i0},k_{iz}), \quad  d^2k_i= dk_{i0} dk_{iz}\quad(i=1,2),
\end{equation}
\begin{equation}
m_2^2=m^2+\vec{k}^2_{2\perp},  \quad m_3^2=m^2+(\vec{p}_{\perp}-\vec{k}_{1\perp}-\vec{k}_{2\perp})^2,
\end{equation}
and the 2-dimensional vector $p'=(p'_0,p'_z)=p-k_1=(p_0-k_{10},p_z-k_{1z})$.

The integral \eqref{Eq:chi} can be computed analytically and for the region $p'^2<(m_2+m_3)^2$ is given by
\begin{equation}
\chi(k_{10},k_{1z};\vec{k}_{1\perp},\vec{k}_{2\perp})=-\frac{i\pi}{p'^2(u_{-}-u_+)}[\log(1-u_-)-\log(-u_-)-\log(-1+u_+)+\log(u_+)], 
\end{equation}
where 
\begin{equation}
u_{\mp}=\frac{1}{2p'^2}\left[p'^2-m_2^2-m^2_3\mp\sqrt{((m_2-m_3)^2-p'^2)((m_2+m_3)^2-p'^2)}\right].
\end{equation}

Similarly, for $p'^2>(m_2+m_3)^2$, we obtain
\begin{equation}
\chi(k_{10},k_{1z};\vec{k}_{1\perp},\vec{k}_{2\perp})=\chi'(k_{10},k_{1z};\vec{k}_{1\perp},\vec{k}_{2\perp})+\chi''(k_{10},k_{1z};\vec{k}_{1\perp},\vec{k}_{2\perp}),
\end{equation}
with
\begin{equation}
\chi'(k_{10},k_{1z};\vec{k}_{1\perp},\vec{k}_{2\perp})=i\pi\frac{\log\frac{m^2_2+m^2_3-p'^2-\sqrt{[p'^2-(m_2-m_3)^2][p'^2-(m_2+m_3)^2]}}{m^2_2+m^2_3-p'^2+\sqrt{[p'^2-(m_2-m_3)^2][p'^2-(m_2+m_3)^2]}}}{\sqrt{[p'^2-(m_2-m_3)^2][p'^2-(m_2+m_3)^2]}},
\end{equation}
and
\begin{equation}
\chi''(k_{10},k_{1z};\vec{k}_{1\perp},\vec{k}_{2\perp})=\frac{2\pi^2}{\sqrt{[p'^2-(m_2-m_3)^2][p'^2-(m_2+m_3)^2]}}.
\end{equation}

The  component of the transverse amplitude, $L_1(\vec{k}_{1{\perp}}, \vec{k}_{2{\perp}})$, can subsequently be written in the form
\begin{equation}
\begin{aligned}
L_1(&\vec{k}_{1{\perp}}, \vec{k}_{2{\perp}})=\\ 
&-i\int_{-\infty}^{\infty}dk_{1z}\left\lbrace \frac{i\pi}{2\tilde{k}_{10}}\left[\chi(\tilde{k}_{10}, k_{1z}; \vec{k}_{1\perp},\vec{k}_{2\perp})v_M(\tilde{k}_{10},k_{1v})+\chi(-\tilde{k}_{10}, k_{1z}; \vec{k}_{1\perp},\vec{k}_{2\perp})v_M(-\tilde{k}_{10},k_{1v})\right]\right.\\
&-\int_0^{\infty}dk_{10}\frac{\chi(-k_{10},k_{1z};\vec{k}_{1\perp},\vec{k}_{2\perp})v_M(-k_{10},k_{1v})-\chi(-\tilde{k}_{10},k_{1z};\vec{k}_{1\perp},\vec{k}_{2\perp})v_M(-\tilde{k}_{10},k_{1v})}{k^2_{10}-\tilde{k}^2_{10}}\\
& \left. -\int_0^{\infty}dk_{10}\frac{\chi(k_{10},k_{1z};\vec{k}_{1\perp},\vec{k}_{2\perp})v_M(k_{10},k_{1v})-\chi(\tilde{k}_{10},k_{1z};\vec{k}_{1\perp},\vec{k}_{2\perp})v_M(\tilde{k}_{10},k_{1v})}{k^2_{10}-\tilde{k}^2_{10}}\right\rbrace,
\end{aligned}
\label{Eq:transverse_final}
\end{equation}
where
\begin{equation}
\tilde{k}_{10}=\sqrt{k^2_{1z}+\vec{k}^2_{1\perp}+m^2}.
\end{equation}
Analogously to the treatment of the BS equation, we have here used subtractions to eliminate the propagator singularities at $k_0=\pm \tilde{k}_{10}$.

\section{Results}\label{Sec:Res}

We multiply the r.h.-side of  Eq.~\eqref{Eq:v} by a factor $\lambda$ and solve this eigenvalue equation by a spline decomposition of the vertex function $v(p,q)$. As inputs we use the scattering length $a$ and the three-body binding energy $B_3$, computed by solving the corresponding problem in Euclidean space \cite{ey3b}. The Minkowski and Euclidean space calculations are then  consistent, as it should be, if the eigenvalue $\lambda=1.0$ is found.

In Table \ref{Tab:results} we show the calculated eigenvalues for three different values of the three-body binding energy, i.e.~$B_3/m=0.006,\: 0.395,\:1.001$. The corresponding values of the scattering length are also listed. It is seen that $\lambda$ acquire a non-zero imaginary part, whereas in all cases the real part of $\lambda$ is very close to unity. For the binding energy $B_3/m=0.395$, the imaginary part is very small (less than 0.2\%). However, for the two other cases the calculation errors result in the imaginary parts about $5\%$ and $10\%$ respectively. It is important to mention that in the Euclidean calculations of Ref.~\cite{ey3b} the full infinite domains of the variables $q_v$ and $q_4$ (and also $k_v$ and $k_4$) were considered by using a mapping procedure. However, due to the many singularities, this is difficult to do in the Minkowski space calculations keeping the same numerical precision, as sizable numbers of basis functions and gauss points are already needed for achieving the aforementioned results. Therefore, following the amplitude decay, we truncated the range of $q_v$ by  $q^{\text{max}}_v/m=6.0$. Similarly, for the binding energies $B_3/m=0.006$ and $0.395$ we used $q^{\text{max}}_0/m=13.0$, whereas  $q^{\text{max}}_0/m=15.0$ for the case $B_3/m=1.001$. In this respect, the calculations are not completely comparable with each other and it partially explains the small non-zero imaginary parts. Naturally, the numerical error is another cause of the imaginary part rise, as well as to the real part not being not exactly one. 
\begin{table}[!htbp]
\centering
\begin{tabular}{c c c}
\toprule
$B_3/m$   & $a m$    &  $\lambda$ \\
\midrule
$0.006$ & $-1.280$  & $0.999-0.054i$ \\
$0.395$ & $-1.500$   & $1.000+0.002i$ \\
$1.001$ & $-1.705$ &  $0.997+0.106i$\\
\bottomrule
\end{tabular}
\caption{Eigenvalues of the three-body ground state  for three scattering lengths, $a$, computed by using the Euclidean three-body binding energies $B_3$ as inputs. \label{Tab:results}}
\end{table}

\begin{figure}[!htbp]
\centering
\includegraphics[scale=0.5]{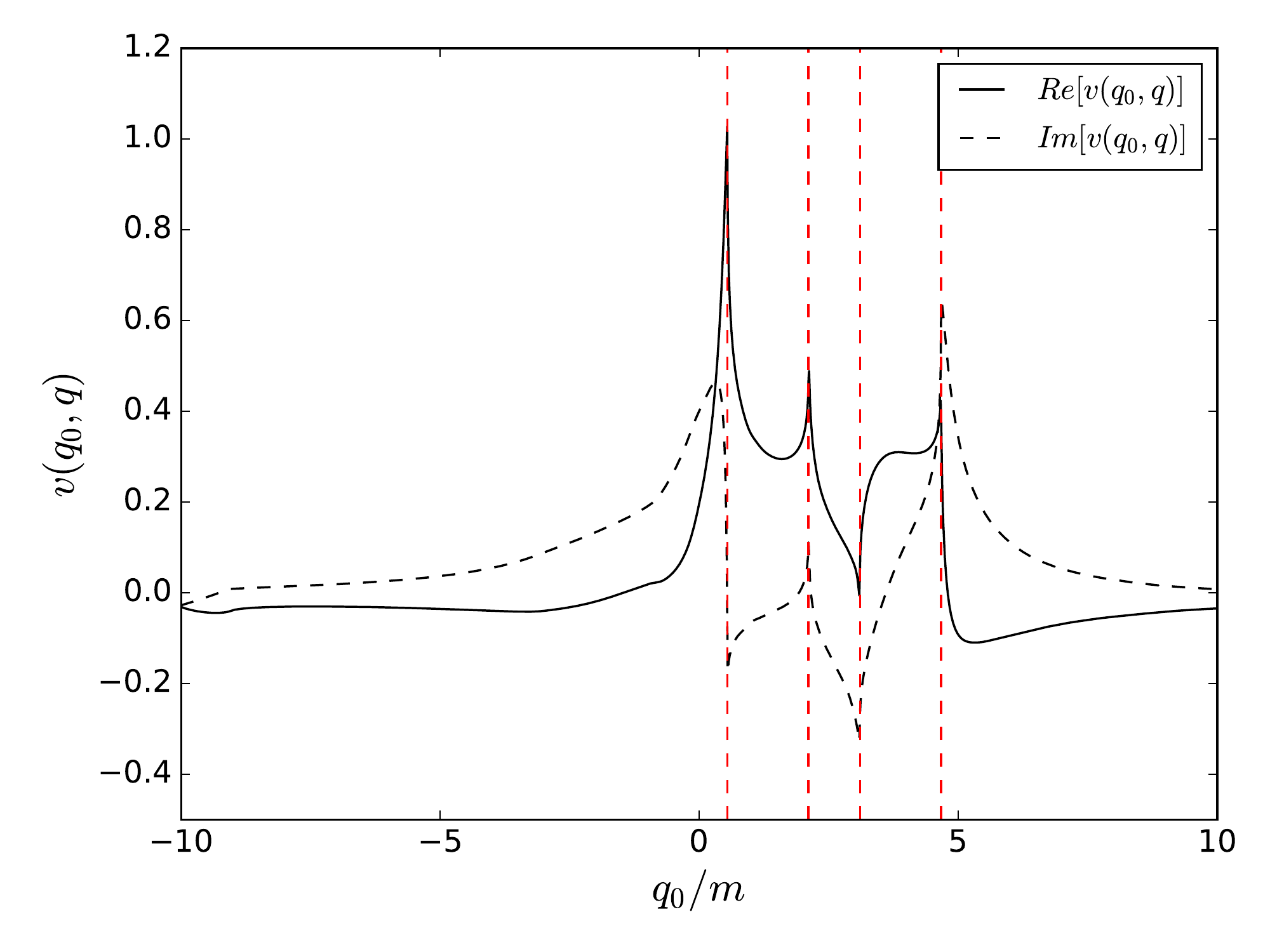} 
\caption{The vertex function, $v(q_0, q_v=0.5 m)$ with respect to $q_0$ for the input parameters $am=-1.5$ and $B_3/m=0.395$. The analytical positions of the peaks, given in Eq.~\eqref{Eq:peaks}, are shown with dashed-red lines.\label{Fig:v}}
\end{figure}

Moreover, in Fig.~\ref{Fig:v} we display the computed vertex function $v(q_0, q_v=0.5 m)$ for the binding energy $B_3/m=0.395$. In the figure we also show the analytical positions of the peaks given by Eq.~\eqref{Eq:peaks} (vertical dashed-red lines), which are matching with the numerical results. These peaks appear in the transition points of the kernel $\Pi(q_0,q_v,\pm \varepsilon_k, k_v)$, 
given by Eqs.  (\ref{Eq:peaks}) in Sec.~\ref{Sec:BSE}. As mentioned in Sec.~\ref{Sec:BSE}, the aforementioned positions  correspond to $M^2_{12}=0$ and $M^2_{12}=4m^2$, i.e.~to the transition points of the two-body scattering amplitude $F(M_{12})$.     

\begin{figure}[!htbp]
\centering
\includegraphics[scale=0.4]{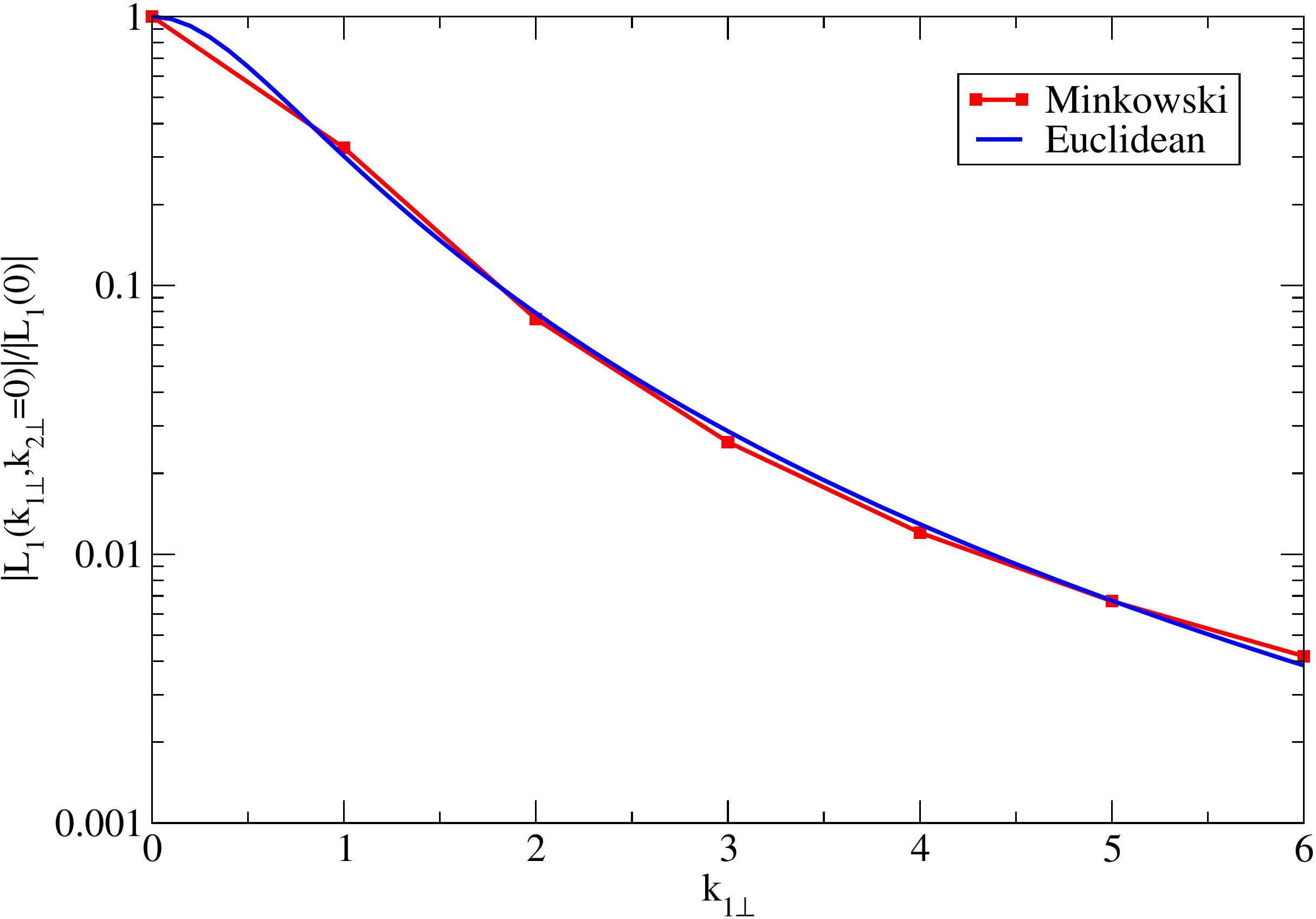}  
\caption{  Transverse amplitude component,  $|L_1(k_{1\perp},k_{2\perp}=0)|$, obtained in Minkowski space compared with the one computed in Euclidean space \cite{ey3b}, for the parameters $am=-1.5$ and $B_3/m=0.395$. \label{Fig:transverse}}
\end{figure}

In Fig.~\ref{Fig:transverse} is shown the modulus of the component $L_1(k_{1\perp},k_{2\perp}=0)$ of the transverse amplitude for $B_3/m=0.395$, computed from Eq.~\eqref{Eq:transverse_final}. We also show, for comparison, the corresponding result by using the solution in Euclidean space. It is seen  that the results are in good agreement with each other. As is clearly visible in Fig.~\ref{Fig:v}, the vertex $v(q_0,q_v)$ v.s.~$q_0$ is a non-smooth function. Despite of this, the obtained transverse amplitude v.s. $k_{\perp}$ is smooth, which makes the coincidence even more remarkable.

\section{Conclusions}\label{Sec:Concl}
We have solved, for the first time, directly in Minkowski space, the three-body BS equation derived in \cite{tobias1} for scalar constituents interacting by the two-body contact interaction. Previously, this equation was solved either using LF projection \cite{tobias1,ck3b}, which attains only the valence component, and, recently, in Euclidean space \cite{ey3b}, which does not give the physical amplitude. The Minkowski BS amplitude is the physical one, which gives direct access to any observable. One needs it to calculate, in particular, the distribution functions of partons in its various forms, and also
 electromagnetic form factors for any  momentum transfer. In the timelike domain any observable requires the Minkowski space solution.

In our method, we reproduced the binding energies found in Ref.~\cite{tobias1,ck3b,ey3b} and also the transverse amplitude found via Euclidean space calculation \cite{ey3b}. This confirms the validity of our approach and correctness of our results. This also brings hope that  solving the three-body BS equation in Minkowski space could be generalized to more realistic interactions and more complex systems, e.g with unequal masses or involving fermions. It is worth to mention that dealing with singular structures numerically can make this method hard to extend for more sophisticated systems. For this reason, we are working on the solution of the three-body equation by means of the Nakanishi integral representation and the LF projection. This latter method was already successfully  used for solving the two-body BS equation, even including spin degrees of freedom, with relatively good numerical stability. Its greatest advantage is the transformation of the initial BS equation into a non-singular integral equation to be solved numerically. This is a work in progress and it is planned for a forthcoming publication.
\bigskip

\textit{Acknowledgements.} We are grateful to Jaume Carbonell for stimulating discussions. 
This study was financed in part by Conselho Nacional de Desenvolvimento Cient\'{i}fico e Tecnol\'{o}gico (CNPq) and 
by Coordena\c{c}\~ao de Aperfei\c{c}oamento de Pessoal de N\'{i}vel Superior - Brasil (CAPES) - Finance code   001.
J.H.A.N. acknowledges the support of the grants \#2014/19094-8 and \#2017/14695-1 and V.A.K. of the grant \#2015/22701-6 from
 Funda\c{c}\~ao de Amparo \`{a} Pesquisa do Estado de S\~ao Paulo (FAPESP). E.Y. thanks for the financial
support of the grant \#2016/25143-7 from FAPESP.  We  thank the FAPESP Thematic Projects
grants   \#13/26258-4 and \#17/05660-0.  V.A.K.~is also sincerely grateful to group of theoretical nuclear physics of ITA, S\~{a}o Jos\'{e} dos Campos, Brazil, for kind hospitality during his visit.

\end{document}